# Hydrogen bonds and asymmetrical heat diffusion in α-Helices. A Computational Analysis.

German A. Miño-Galaz[*,1,2,3] and Gonzalo Gutierrez[1]

[1]Group of NanoMaterials (www.gnm.cl), Departamento de Física, Facultad de Ciencias, Universidad de Chile, Casilla 653, Santiago, Chile

[2]Centro Interdisciplinario de Neurociencias de Valparaíso (CINV), Universidad de Valparaíso, Valparaíso, Chile

[3]Facultad de Ciencias Biologicas, Centro de Bioinformatica y Biologia Integrativa, Universidad Andres Bello, Av.Republica 239, Santiago, Chile

*Corresponding author: germino@u.uchile.cl

**Abstract.**

In this work, we report the heat rectifying capability of α-helices. Using molecular dynamics simulations we show an increased thermal diffusivity in the C-Terminal to N-Terminal direction of propagation. The origin of this effect seems to be a function of the particular orientation of the hydrogen bonds stabilizing these α-helices. Our results may be relevant for the design of thermal rectification devices for materials science and lend support to the role of normal length hydrogen bonds in the asymmetrical energy flow in proteins.

## Introduction.

Hydrogen bonds have been an inexhaustible source of research for decades (1). Their role as stabilizing agents of protein structures has been clearly established (2). These also modulate the chemical reactivity of enzymes through changes in the physicochemical properties of the surrounding structural elements (3-5). Likewise, the role of hydrogen bonds in thermal conduction in proteins has recently been reported, revealing their importance in heat diffusion across α-helices (6) and the β-sheet structure of spider silk (7).

Proteins have been naturally selected under evolutionary pressures, and are therefore an interesting source of molecular examples for thermal control. Phenomena such as conformational changes, enzyme catalysis, allosteric cooperativity, and intermolecular affinities, among other processes (8), require of a high degree of control over thermal energy flow. In this respect, nonlinear excitations such as breathers (9), solitons (10-12), and low and high energy vibrational modes (8,13), have been drawing increased attention as mediators of energy transport in biomolecules. There have been significant efforts to study energy flow in proteins and the relation of this flow to protein dynamics (14-18). Research on heat flow in different structures, such as peptide helices (6,16-17,19-20), heme cofactors (18), beta sheets(7), and functionalized materials (21-22), have been reported in literature. These reports have shown that excess energy deposited at particular sites can propagate along structures through the covalent backbone of the molecules (17,22) and through weaker interactions such as hydrogen bonds (6,7,21,26).

Heat flow in protein structures has been recently correlated to the known pathways for allosteric communication. This means that heat propagates preferentially along the same structural pathway that defines the allosteric communication route in proteins (23-26). Allosteric communication is an unsolved problem in biochemistry that determines the processes of signal transmission through proteins on both short (3 Å) and long (100 Å) range scales. This phenomenon is involved in crucial cellular and physiological functions and is a determinant for serious human diseases (27-29) Allosteric communication can be determined using both experimental (27-29) and theoretical (23-26) techniques, and its occurrence has been clearly established. Thermodynamically, the free energy of allosteric processes in proteins is interpreted by attributing the enthalpic part (ΔH) to conformational changes in the protein structure and entropic contribution (ΔS) to the vibrational properties of a protein (30-32). Based on this, three types of allosteric signaling are distinguished: Type I, with small or subtle conformational changes governed largely by entropy, (ΔH≈0); Type II, with the participation of enthalpic and entropic components to different extents; and Type III, largely governed by enthalpic changes (ΔS≈0) (30). Thus, Type I allosteric communication is mediated by vibrational modes of the proteins that can act as energy carriers (30-31,33). In particular, hydrogen bonds have been recently proposed as an important element in the energy flux associated with Type I allosteric communication in the PDZ-2 protein, which can, in turn, transfer heat with preferential directionality (26). Thus, hydrogen bonds may be an important supporting structure for signal propagation in proteins and in determining the expected directionality of a signal that propagates from the allosteric regulatory site to the effector site.

Understanding heat flow at the molecular and nanoscale levels is currently of high interest in the development of phononics (34-36). Control of heat flow in materials science is highly desirable for implementing thermal devices such thermal logic gates (37); thermal memories (38), and acoustic and thermal cloakers (34). Fundamental for the implementation of these phononic devices is the thermal diode that rectifies thermal energy passage in a predetermined direction (34,39). Experimental heat flow rectification has been achieved using mass-graded carbon and boron nitrite nanotubes (40), as well as with vanadium dioxide composites (41) in which a symmetry-breaking structure joins the thermal source with the thermal drain. The rectifying capability of symmetry-breaking structures has also been computationally demonstrated in carbon nanotube intramolecular junctions (42), carbonano-cones (43), and in asymmetric graphene ribbons (44).

In this work, we report the rectifying capability of a set of α-helices, that is, the capability of heat transfer with preferential directionality. Using molecular dynamics simulations, energy flux was studied by vibrational excitation on the N- or C-terminal sides of the α-helices, independently (Fig. 1). Increased thermal diffusivity was found in the C-Terminal to N-Terminal direction of heat diffusion, showing that α-helices may act as thermal rectifiers. Heat flow in both directions of propagation was analyzed in terms of the global temperature for each structure; average power; and the kinetic and potential energy of the atoms and bonds involved in selected hydrogen bonds. All of the results suggest that backbone hydrogen bonds, and their natural asymmetry (-C=O···H–N-), are a symmetry-breaking element that gives rise to the observed thermal rectification. The presented results are consistent with computational reports about hydrogen-bond mediated directional thermal energy transport in functionalized hydrophobic and hydrophilic silica−water interfaces (21). Moreover, the results lend support to a recent report suggesting that normal length hydrogen bonds act as thermal rectifiers in proteins, as well as supporting the possible role of these bonds in the selective and directional propagation of allosteric signals in proteins (26).

**Methods and Computational Details.**

Heat diffusion evaluations were performed for five α-helical structures. Each α-helix was composed by only one type of amino acid residue, and each had a length of 33 residues. The group of five α-helices was constituted by phenilalanine (PHE), isoleucine (ILE), leucine (LEU), methionine (MET), and valine (VAL). The α-helices were acetylated and amidated at their N- and C-terminals, respectively, thus rendering them electrically neutral.

Figure 1 shows the structure of poly-isoleucine, with hydrogen bonds and the backbone structure highlighted. The structures were minimized over $10^5$ steps and further equilibrated *in vacuo* for 1 ns at a constant temperature of 300 K using Langevin dynamics with a damping coefficient of $\gamma = 5$ ps$^{-1}$. An integration time-step of 1 fs was used with a uniform dielectric constant of 1 and a cutoff of non-bonded forces, with a switching function that started at a distance of 10 Å and reached zero at 13.5 Å. All of the selected systems kept an $\alpha$-helical structure without the need for any specific constraints. After this, all the systems were cooled and re-equilibrated to 10 K for 1 ns.

The heating procedure was then applied with this initial temperature of 10 K. Heat was then injected by increasing the kinetic energy of the carbonyl group (C=O) located at residue 1 of each structure (Fig. 1) at the N-acetyl moiety to a target temperature of 300 K. With this point of thermalization, vibrational energy flow occurred from the N-Terminal to C-Terminal direction, termed in this work as the "direct" or "N-Term → C-Term" direction of propagation. For this, the same C=O group was subjected to a confining potential of 5 kcal/mol*A$^2$, and two specific atoms were fixed. These fixed atoms were located in the N-acetylated moiety (carbon atom of the methyl group) and in the C-amidated moiety (nitrogen atom of the amide group). These two atoms, in combination with the confining potential of the C=O group, prevented the loss of the helical structure during the heating process. These conditions also avoided the dissipation of energy by internal rotation along the axis connecting the fixed atoms in the N- and C-terminals.

A second heat procedure consisted in increasing the kinetic energy of the N-H group, located at residue 33 of each structure, to a target temperature of 300 K. With this point of thermalization, vibrational energy flow occurred from the C-Terminal to the N-Terminal direction, termed in this work as the "inverse" or "N-Term ← C-Term" direction of propagation. The same N-H group was subjected to a confining potential of 5 kcal/mol*A$^2$, and two specific atoms were fixed. These fixed atoms were located in the N-acetylated moiety (carbon atom of the methyl group) and in residue 33 ($\alpha$ carbon atom). The inclusion of confining potentials and fixed atoms followed the same reasoning explained above. Hydrogen atoms dynamics was maintained free along the simulations by adding the option rigidbonds=none.

The temperature of the each structure was determined for the direct and inverse directions as a function of time. Structurally, the formation of total hydrogen bonds was measured alongside simulations for both directions of propagation. Heat flow was determined in selected atoms through a running average of 250 or 1000 time-steps. With this data, kinetic, potential energy, and average power were computed for selected atoms. All simulations were performed *in vacuo.* There is experimental evidence that 70% of the injected energy in peptide helices dissipates in the solvent(17). Thus, since the interest was to study heat flow through the structure, simulations took place *in vacuo* to maximize this process. This model setup sought to emulate a non-fluxtional α-helix and its response to local energy excitation.

All simulations were performed with the NAMD Molecular Dynamics Software v.2.9 (45) using the CHARMM22 (46) potential energy function with CMAP correction. CMAP is an energy correction based on quantum calculations that improves protein backbone behavior and thus yields more accurate predictions of protein structure and dynamics (47). Heating procedures used the T-couple algorithm of NAMD2. Visualization and data analyses were carried out using the VMD 1.9.1 software (48).

**Results and Discussion.**

Figure 2 shows temperature as a function of time for each structure, both for direct (N-Term → C-Term) and inverse (N-Term ← C-Term) directions. Insets in Figure 2 show the formation of total hydrogen bonds alongside both directions of propagation. These plots give a general picture of the thermal diffusivity found along each structure in the direct and inverse directions. Moreover, inspection of Figure 2 clearly shows differences between the two directions of heat diffusion. Faster thermalization was observed for the inverse direction whereas slower thermalization was observed for the direct direction. Given this, inverse propagation more efficiently thermalized the α-helices. The quantified number of hydrogen bonds was slightly higher for the direct direction in the cases of VAL, PHE, LEU, and ILE. Only in the case of MET was the number of hydrogen bonds similar along both directions. These first observations suggests the particular orientation of hydrogen bonds that stabilize the α-helices (N-Term-C=O···H–N-C-Term, in all cases), and this orientation was also the cause for the increased diffusivity observed for the inverse, or N-Term ← C-Term, direction.

In order to gain numerical insight into the heat diffusion process along both directions, the average power was determined for the backbone atoms N, O, C, and Cα at residue 28 for the direct direction and at the backbone atoms N, O, C, and Cα at residue 6 for the inverse direction. The results of this are shown in Table 1. The average power computed in symmetrically opposite positions for both directions were between 3 - 8 units of difference, favoring the inverse direction. These comparisons suggest that the natural asymmetry of the hydrogen bonds may result in these bonds acting as thermal rectifying devices in α-helices.

To further evaluate this idea, the kinetic energy of N, H, and O atoms of selected hydrogen bonds in each structure was estimated in both flow directions. In particular, analyses were performed for the hydrogen bonds at positions 14 – 18 for MET and PHE; positions 10 – 14 for VAL and LEU; and positions 13 – 17 for ILE. The analyses showed consistent and significant differences in both directions of propagation for all systems (Fig. 3). In the direct direction, the kinetic energy of O and N split, showing increased energy for the O atom. For the inverse direction, the kinetic energy of O and N evolved together, keeping similar values as energy propagation took place. The increase in energy observed for the O atom in the direct direction suggests that the energy tended to get trapped when injected into the N-terminal side of the α-helices, while for the inverse direction, the energy freely flowed along the α-helices. These results imply that for the particular orientation of hydrogen bonds in α-helices, heat diffuses easily in the inverse direction (C=O ← H–N) and with difficultly in the direct direction (C=O → H–N). The origins of this effect may be due to the structural asymmetry of the hydrogen bonds and to the consequent stretching force constants associated with the N-H and C=O bonds (Fig. 4). For N-H and C=O, the stretching force constant ($K_b$) is 440 kcal·mol$^{-1}$·Å$^{-2}$ and 620 kcal·mol$^{-1}$·Å$^{-2}$, respectively. Thus, behavior on both sides of the hydrogen bond was not the same, with the C=O side showing 620/440 = 1.4 times higher ability to trap energy in its vibration, taking the ratio as a rough index. The heat injected into the N-terminal side (direct direction) first reached the C=O moiety of each hydrogen bond. This point seems to act as a kind of "hard" end that trapped heat. In the inverse direction, when the heat was injected into the C-terminal side, it first reached the H–N moiety of each hydrogen bond. This moiety seemed to act as a kind of "soft" end, and it was less effective than its counterpart (C=O) in trapping heat. This effect was multiplied by the c.a. 30 hydrogen bonds per structure that connected in a series and gave rise to the overall thermal rectification observed in the simulations.

To further verify the observations described above, potential energy analyses was performed on the same hydrogen bonds as in kinetic energy analyses, that is, positions 14 - 18 for the MET and PHE; positions 10 – 14 for VAL and LEU; and positions 13 – 17 for ILE. For potential energy analyses the Hooke law was used, V(bond)= $K_b (x-x_0)^2$, where $K_b$ is the spring constant of each bond and $x_0$ is the equilibrium distance, which is equal to 0.997 Å and 1.230 Å for N-H and C=O bonds, respectively (½ pre-factor of Hooke law is included in Kb). The results of these analyses are shown in Figure 5. Similar trends and significant differences were observed in both directions of propagation for all systems. The direct direction consistently showed higher potential energy values in the C=O bond, while the inverse direction showed lower potential energy in the C=O bond. The increase in potential energy observed for the C=O bond in the direct direction indicates that energy tends to get trapped here as it is injected in the N-terminal side of the α-helices, while for the inverse direction, energy can freely flow along the α-helices. Therefore, the same interpretation offered for kinetic energy analyses holds here; that is, the C=O moiety appears to act as a "hard" end that traps heat while the H–N moiety acts as a "soft" end.

Based on these observations, we specifically propose that normal length hydrogen bonds can act as a thermal rectifier, or, in other words, can act as a structure that transfers thermal energy with preferential directionality.

**Conclusions.**

Motivated by a previous report about the role of hydrogen bonds in heat diffusion in α-helices[6] and by the observations of Schoen et al. (2009) about a possible rectifying role of hydrogen bonds in functionalized materials, we decided to investigate the effects of applying heat to a set of α-helices at the opposite side than that previously reported. The particular set of α-helices used in these previous and this present study are stabilized by c.a. 30 hydrogen bonds connected in a series along the helix backbone. This characteristic makes them very suitable models to study the effects of hydrogen bond orientation in structure thermalization in the "direct" (N-term → C-Term) and "inverse" (N-Term ← C-Term) directions.

The results showed that the inverse direction had faster thermalization while the direct direction had slower thermalization. Thermalization rates were also followed by measuring the average power at the O, N, C, and Cα atoms of residues 28 and 6 for the direct and inverse heating procedures, respectively. Kinetic and potential energy analyses of selected hydrogen bonds in both directions suggested that these act as the underlying chemical motif for thermal rectification. It seems that the thermal mechanism follows the intrinsic asymmetry of the hydrogen bonds, -C=O···H–N-. When the heat injected into the N-terminal side (direct direction) first reached the C=O moiety of each hydrogen bond, this point acted as a kind of "hard" end that trapped heat, which was reflected by the increased kinetic and potential energy of this bond. In the inverse direction, when heat was injected into the C-terminal side, it first reached the H–N moiety of each hydrogen bond. This moiety seemed to act as a kind of "soft" end that was less effective than its counterpart (C=O moiety) at trapping heat. Due to its lower spring constant, the H–N moiety could not trap energy in the same way as the C=O moiety. The overall effect was that heat diffused more efficiently than in the inverse situation, and the favored heat diffusion direction of C=O ← H–N held.

We conclude that the preferential direction of heat diffusion is directed by the particular orientation of hydrogen bonds. In $\alpha$-helices, this orientation of hydrogen bonds is conserved along the structure, and the effect is therefore multiplied by the number of hydrogen bonds that support the helical structure. Our findings are consistent with the proposal of Schoen et al. (2009), in which hydrogen bonds may act as thermal diodes. Recent evidence has shown a correlation between the heat diffusion pathways and the known allosteric communication pathways in proteins (23-26). The present results support the role of hydrogen bonds in heat diffusion in proteins (7-6), the function of normal length hydrogen bonds as thermal rectifiers in proteins and that these bonds can operate a source of directional energy flow in proteins (26). The study and understanding of heat flow directionality in materials – phonon rectification – is desirable in materials science for the development of thermal gates (37), thermal memories (38), and thermal cloakers (34). In this respect, the inclusion of $\alpha$-helices in the designs of materials science may be a useful alternative for implementing these thermal devices.

**References**

1. Pimentel, G. C.; McClellan, L. The Hydrogen Bond; W. H. Freeman and Company: San Francisco and London, 1960.

2. Nick Pace, C.; Scholtz, J. M.; Grimsley, G. R. Forces stabilizing proteins. *FEBS Lett.* 2014, 588, 2177−2184.

3. Frey, P. A.; Whitt, S. A.; Tobin, J. B. A low-barrier hydrogen bond in the catalytic triad of serine proteases. *Science* 1994, 264, 1927−30.

4. Miño, G.; Contreras, R. On the role of short and strong hydrogen bonds on the mechanism of action of a model chymotrypsine active site. *J. Phys. Chem. A* 2009, 113, 5769−5772.

5. Mino, G.; Contreras, R. Non-electrostatic components of short and strong hydrogen bonds induced by compression inside fullerenes. *Chem. Phys. Lett.* 2010, 486, 119−122.

6. Miño, G.; Barriga, R.; Gutierrez, G. Hydrogen bonds and heat diffusion in α-helices: a computational study. *J. Phys. Chem. B.* 2014, 118, 10025-34.

7. Zhang L, Chen T,Bana H, Liu L Hydrogen bonding-assisted thermal conduction in β-sheet crystals of spider silk protein. *Nanoscale* 2014, 6, 7786-7791.

8. Leitner, D. M.; Straub, J. E., Eds. Proteins: Energy, Heat and Signal Flow; Taylor and Francis Group: New York, 2010.

9. Piazza, F.; Sanejouand, Y. H. Discrete breathers in protein structures. *Phys. Biol.* 2008, 5, 026001.

10. Davydov, A. S. Solitons and energy transfer along protein molecules. *J. Theor. Biol.* 1977, 66, 379−387.

11. Cruzeiro-Hansson, L.; Takeno, S. Davydov model: The quantum, mixed quantum-classical, and full classical Systems. *Phys. Rev. E* 1997, 56, 894.

12. Mimshe Fewu, J. C.; Tabi, C. B.; Edongue, H.; Ekobena Fouda, H. P.; Kofané, T. C. Wave patterns in α-helix proteins with interspine coupling. *Phys. Scr.* 2013, 87, 025801.

13. Leitner, D. M. Energy flow in proteins. *Annu. Rev. Phys. Chem.* 2008, 59, 233−259

14. Lervik, A.; Bresme, F.; Kjelstrup, S.; Bedeaux, D.; Miguel Rubi, J. Heat transfer in protein-water interfaces. *Phys. Chem. Chem. Phys.* 2010, 12, 1610−7.

15. Erman, B. Relationship between ligand binding sites, protein architecture and correlated paths of energy and conformational fluctuations. Phys. Biol. 2011, 8, 0560003.

16. Helbing, J.; Devereux, M.; Nienhaus, K.; Nienhaus, G. U.; Hamm, P.; Meuwly, M. Temperature dependence of the heat diffusivity of proteins. *J. Phys. Chem. A* 2012, 116, 2620−8.

17. Botan, V.; Backus, E. H.; Pfister, R.; Moretto, A.; Crisma, M.; Toniolo, C.; Nguyen, P. H.; Stock, G.; Hamm, P. Energy transport in peptide helices. *Proc. Natl. Acad. Sci. U. S. A.* 2007, 104, 12749−12754.

18. Sagnella, D. E.; Straub, J. E. Directed Energy "Funneling" Mechanism for Heme Cooling Following Ligand Photolysis or Direct Excitation in Solvated Carbonmonoxy Myoglobin. *J. Phys. Chem. B* 2001, 105, 7057−7063.

19. Schade, M.; Moretto, A.; Crisma, M.; Toniolo, C.; Hamm, P. Vibrational Energy Transport in Peptide Helices after Excitation of C− D Modes in Leu-d 10. *J. Phys. Chem. B* 2009, 113, 13393-13397.

20. Backus, E. H.; Nguyen, P. H.; Botan, V.; Pfister, R.; Moretto, A.; Crisma, M.; Toniolo, C.; Stock, G.; Hamm, P. Energy transport in peptide helices: A comparison between high-and low-energy excitations. *J. Phys. Chem. B* 2008, 112, 9091-9099.

21. Schoen, P. A. E.; Michelb, B.; Curionib, A.; Poulikakosa, D. Hydrogen-bond enhanced thermal energy transport at functionalized, hydrophobic and hydrophilic silica−water interfaces. *Chem. Phys. Lett.* 2009, 476, 271−276.

22. Lin, Z.; Rubtsov, I. V. Constant-speed vibrational signaling along polyethyleneglycol chain up to 60-Å distance. *Proc. Natl. Acad. Sci. U. S. A.* 2012, 109, 1413−1418.

23. Ota, N.; Agard, D. A. Intramolecular signaling pathways revealed by modeling anisotropic thermal diffusion. *J. Mol. Biol*. 2005, 351, 345−354

24. Burendahl, S.; Nilsson, L. Computational studies of LXR molecular interactions reveal an allosteric communication pathway. *Proteins* 2012, 80, 294−306.

25. Liu, J.; Tawa, G. J.; Wallqvist, A. Identifying cytochrome p450 functional networks and their allosteric regulatory elements. *PLoS One* 2013, 8, e81980.

26. Miño-Galaz, G.A. Allosteric Communication Pathways and Thermal Rectification in PDZ-2 Protein: A Computational Study. 2015 J. Phys. Chem. B,J. Phys. Chem. B, Article ASAP DOI: 10.1021/acs.jpcb.5b02228 Publication Date (Web): May 1, 2015.

27. Laine, E.; Auclair, C.; Tchertanov, L. Allosteric communication across the native and mutated KIT receptor tyrosine kinase. *PloS Comput. Biol.* 2012, 8, e1002661.

28. Noinaj, N.; Bhasin, S. K.; Song, E. S.; Scoggin, K. E.; Juliano, M. A.; Juliano, L.; Hersh, L. B.; Rodgers, D. W. Identification of the allosteric regulatory site of insulysin. *PLoS One* 2011, 6, e20864.

29. Seldeen, K. L.; Deegan, B. J.; Bhat, V.; Mikles, D. C.; McDonald, C. B.; Farooq, A. Energetic coupling along an allosteric communication channel drives the binding of Jun-Fos heterodimeric transcription factor to DNA. *FEBS J.* 2011, 278, 2090−104.

30. Tsai, C.J.; Del Sol, A.; Nussinov, R. Allostery: absence of a change in shape does not imply that allostery is not at play. J Mol Biol, 2008, 378,1-11

31. Cooper, A.; Dryden, D.T. Allostery without conformational change. A plausible model. *Eur Biophys J.*, 1984, 11, 103-9.

32. Monod, J.; Wyman, J.; Changeux, J. P. On the nature of allosteric transitions: A plausible model. *J. Mol. Biol.* 1965, 12, 88−118.

33. Fuentes, E. J.; Gilmore, S. A.; Mauldin, R. V.; Lee, A. L. Evaluation of energetic and dynamic coupling networks in a PDZ domain protein. *J. Mol. Biol.* 2006, 364, 337−51.

34. Maldovan, M. Sound and heat revolutions in phononics. *Nature* 2013, 503, 209–217.

35. Li, N.; Ren, J.; Wang, L.; Zhang, G.; Hänggi, P.; Li, B. Colloquium: Phononics: Manipulating heat flow with electronic analogs and beyond. *Rev. Mod. Phys.* 2012, 84, 1045.

36. Wang, L.; Li, B. Phononics gets hot. *Phys. World.* 2008, 21, 27-29.

37. Wang, L.; Li, B. Thermal logic gates: computation with phonons. *Phys. Rev. Lett.* 2007 ,99, 177208.

38. Wang, L.; Li, B. Thermal memory: a storage of phononic information. *Phys. Rev. Lett.* 2008, 101, 267203.

39. Li, B., Wang, L., & Casati, G. Thermal diode: rectification of heat flux. *Phys. Rev. Lett.* 2004, 93, 184301.

40. Chang, C. W.; Okawa, D.; Majumdar, A.;Zettl, A. Solid-state thermal rectifier. *Science*, 2006, 314, 1121-1124.

41. Zhu, J.; Hippalgaonkar, K.; Shen, S.; Wang, K.; Abate, Y.; Lee, S.; Wu, J.; Yin, X.; Majumdar, A.; Zhang, X. Temperature-Gated Thermal Rectifier for Active Heat Flow Control. *Nanoletters* 2014 14, 4867-4872.

42. Wu, G., & Li, B. Thermal rectification in carbon nanotube intramolecular junctions: Molecular dynamics calculations. *Phys. Rev. B* 2007, 76, 085424.

43. Yang, N., Zhang, G., & Li, B.  Carbon nanocone: a promising thermal rectifier. *Appl. Phys. Lett*. 2008, 93, 243111.

44. Yang, N., Zhang, G.; Li, B. Thermal rectification in asymmetric graphene ribbons. *Appl. Phys. Lett*. 2009, 95, 033107.

45. Kale, L.; Skeel, R.; Bhandarkar, M.; Brunner, R.; Gursoy, A.;  Krawetz, N.; Phillips, J.; Shinozaki, A.; Varadarajan, K.; Schulten, K.  NAMD2: Greater scalability for parallel molecular dynamics. *J. Comput. Phys.* 1999, 151, 283−312.

46. Mackerell, A. D., Jr.; Feig, M.; Brooks, C. L., 3rd. Extending the  treatment of backbone energetics in protein force fields: limitations of gas-phase quantum mechanics in reproducing protein conformational distributions in molecular dynamics simulations. *J. Comput. Chem.* 2004, 25, 1400−1415

47. MacKerell, A. D., Jr; Bashford, D.; Bellott, M.; Dunbrack, R. L., Jr.; Evanseck, J.; Field, M. J.; Fischer, S.; Gao, J.; Guo, H.; Ha, S.; et al. All-atom empirical potential for molecular modeling and dynamics studies of proteins. *J. Phys. Chem. B* 1998, 102, 3586−3616.

48. Humphrey W, Dalke A, Schulten K  VMD—visual molecular dynamics. *J. Mol. Graph*. 1996, 14, 33–38

**Acknowledgements.**

The authors would like to acknowledge funding provided by FONDECYT Grant 3110149 (awarded to G. Miño) and the partial support of FONDECYT Regular 1120603 (awarded to G. Gutierrez). This work was also partially supported by grants CONICYT Proyecto Anillo ACT – 1104, FONDECYT 1131003 and Millennium Initiative P09-022-F.

**Table and Figure Legends**

**Table 1.** Name, structure, and average power in the 40 – 100 ps for the backbone atoms O, N, C, and Cα of residues 28 and 6 for the N-Term → C-Term and N-Term ← C-Term procedures, respectively.

| Name and structure | Average power (pW) at O, N, C, and Cα atoms of residue 28 N-Term → C-Term | Average power (pW) at O, N, C, and Cα atoms of residue 6 N-Term ← C-Term |
|---|---|---|
| MET 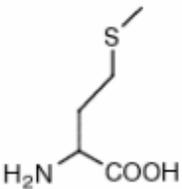 | 20.12 | 26.14 |
| VAL 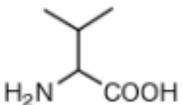 | 19.31 | 27.43 |
| PHE 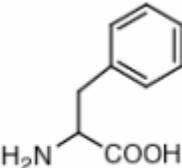 | 21.20 | 29.40 |
| LEU 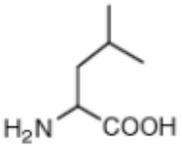 | 19.23 | 22.24 |
| ILE 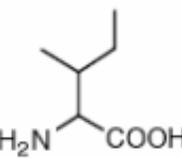 | 24.57 | 27.33 |

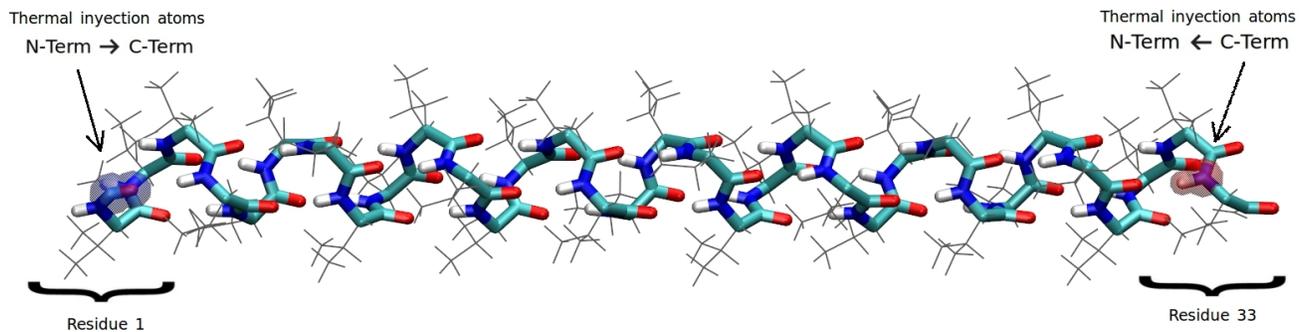

**Figure 1.** Model for poly-isoleucine showing the thermal injection atoms for N-Term → C-Term propagation at residue 1 and for N-Term ← C-Term propagation at residue 33.

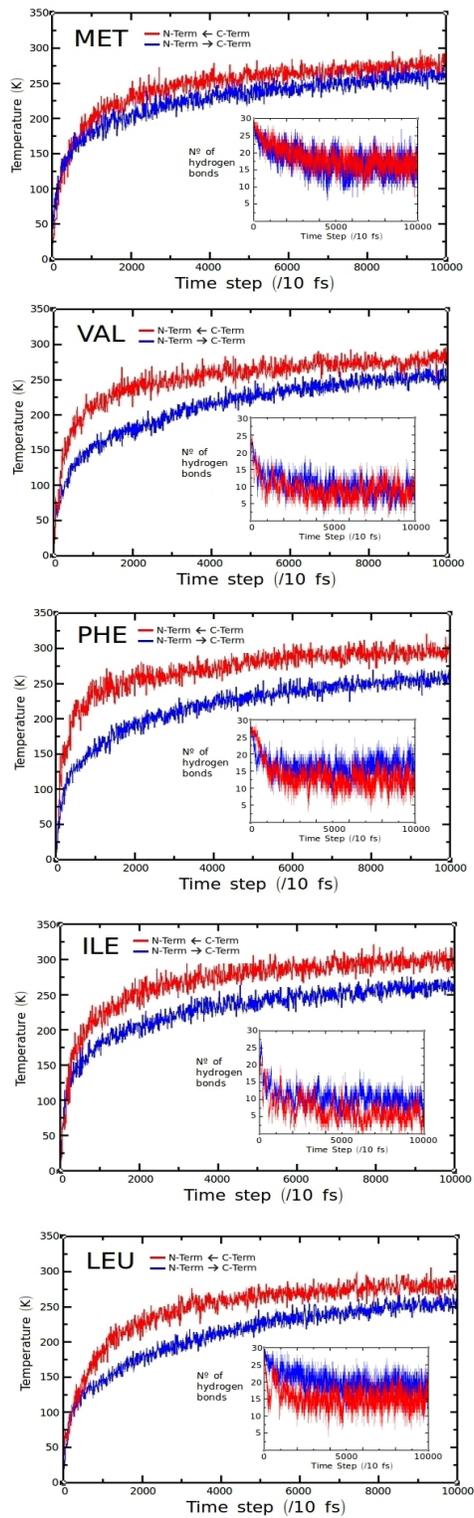

**Figure 2.** Heating curves for N-Term → C-Term propagation in blue and for N-Term ← C-Term propagation in red. Insets represents hydrogen bonds formed in each structure during the heating procedure, with N-Term → C-Term in blue and N-Term ← C-Term in red.

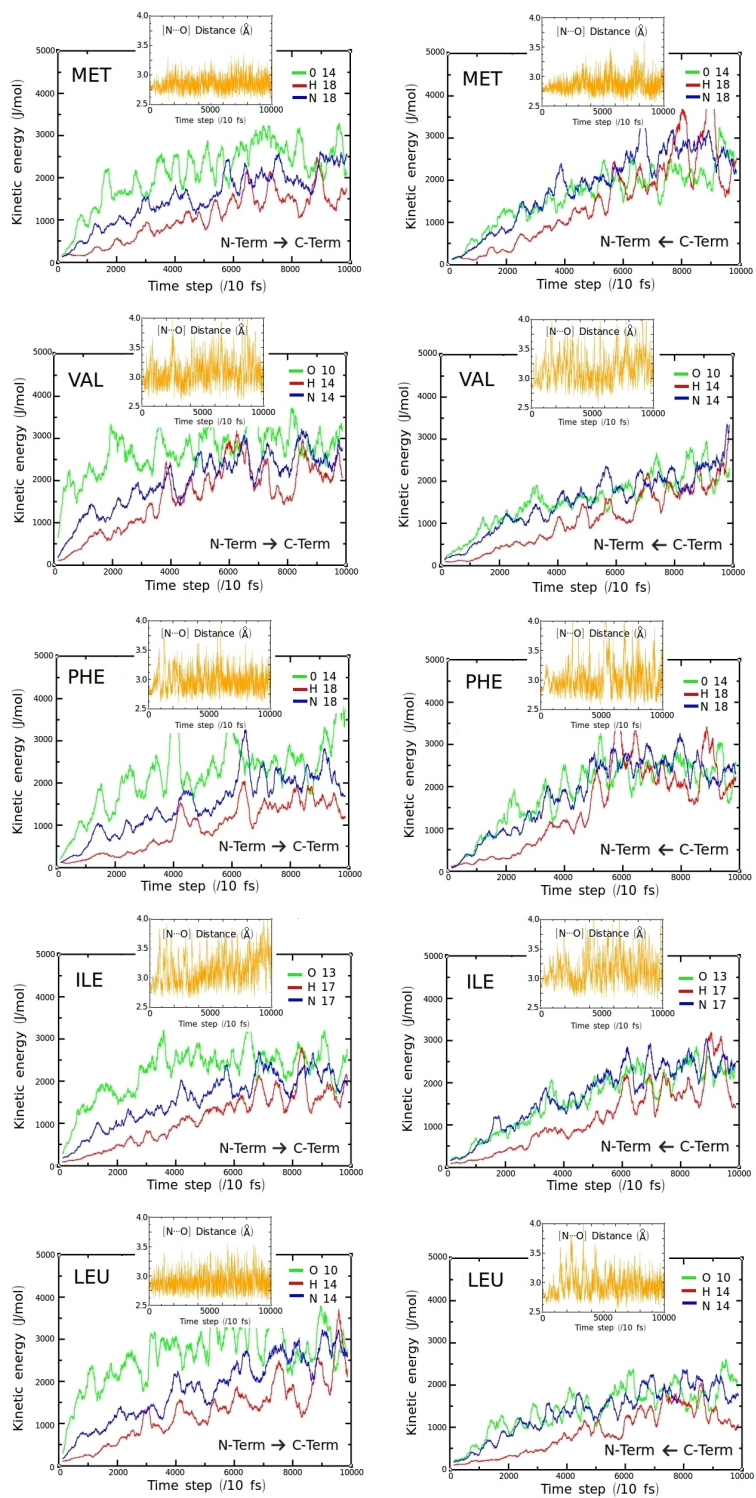

**Figure 3.** Plots of kinetic energy versus time measured at the atoms of selected hydrogen bonds for each structure. Left column represents N-Term → C-Term propagation and right column represents N-Term ← C-Term propagation. Insets represent the [N⋯O] distance for the analyzed hydrogen bond for each structure.

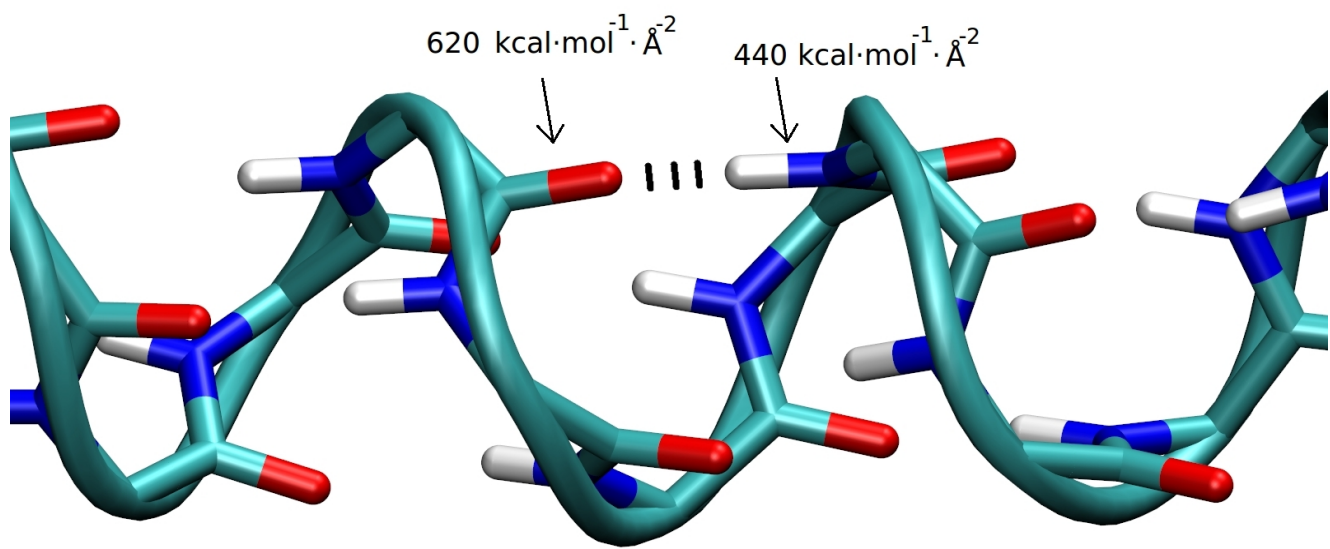

**Figure 4.** Backbone of a generic α-helix structure that represents the particular orientation of the hydrogen bonds and the respective force constant for the C=O and N-H moieties.

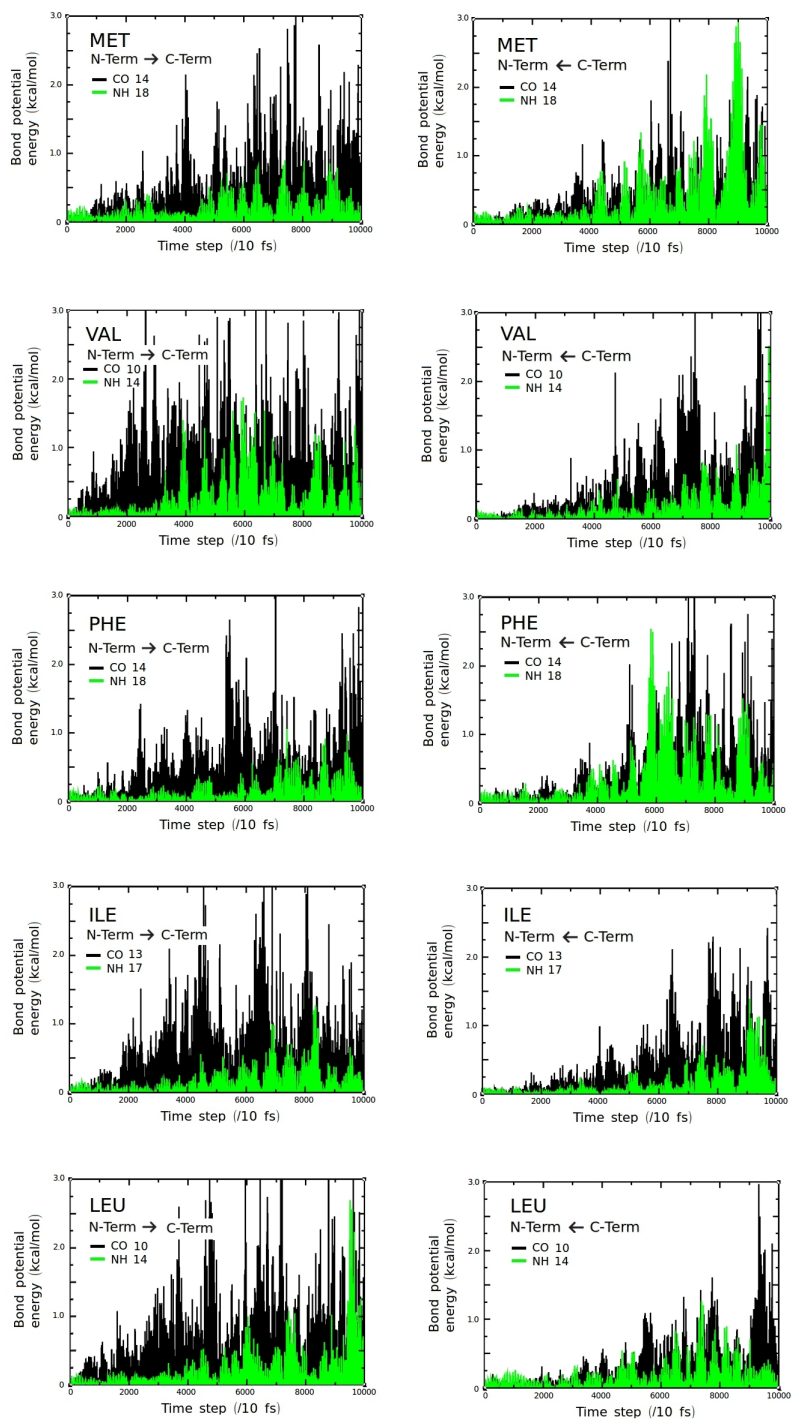

**Figure 5.** Potential bond energy versus time measured at selected hydrogen bonds for each structure. Left column represents direct or N-Term → C-Term propagation and right column represents inverse or N-Term ← C-Term propagation.

**Graphical Abstract**

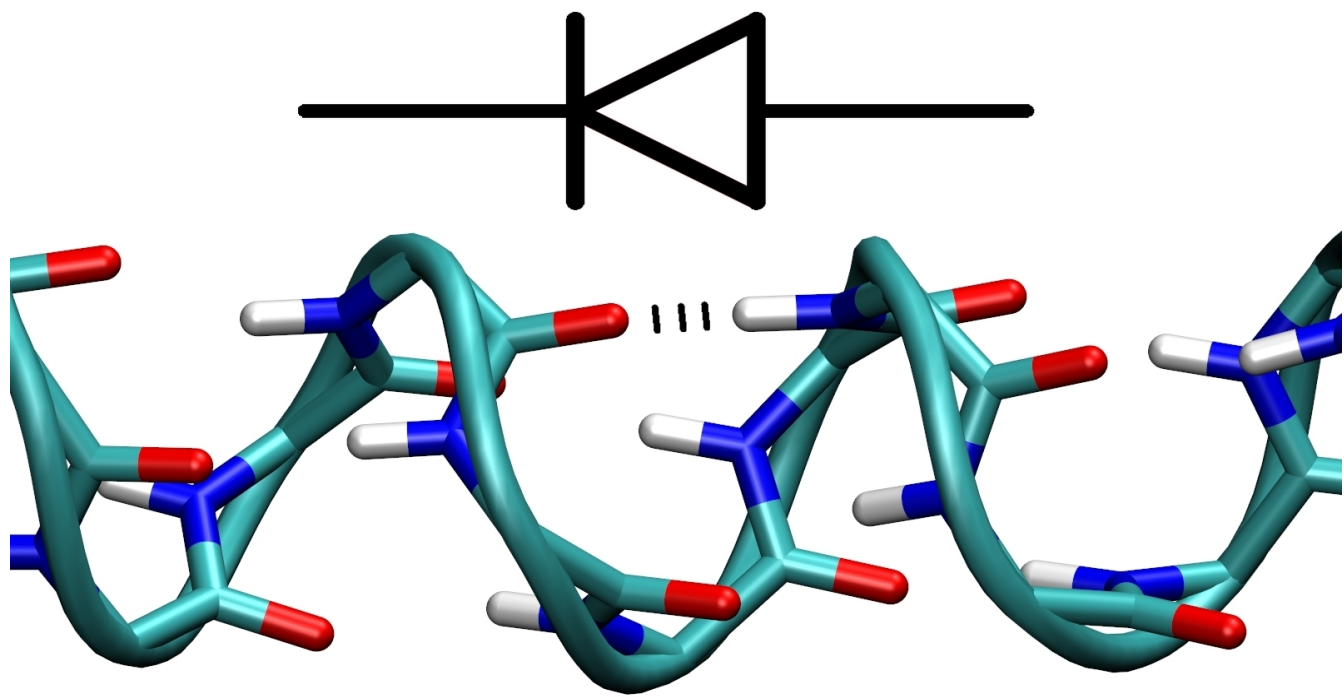

Heat rectification through hydrogen bonds